# Mid-infrared VIPA Spectrometer for Rapid and Broadband Trace Gas Detection

Lora Nugent-Glandorf,[1,*] Tyler Neely,[1] Florian Adler,[1] Adam J. Fleisher,[2] Kevin C. Cossel,[2] Bryce Bjork,[2] Tim Dinneen,[3] Jun Ye,[2] and Scott A. Diddams[1,*]

[1]*National Institute of Standards and Technology, Time & Frequency Division, 325 Broadway, Boulder, CO 80305*
[2]*JILA, National Institute of Standards and Technology and University of Colorado, 440 UCB, Boulder, CO 80309*
[3]*Precision Photonics, 3180 Sterling Circle, Boulder, CO 80301*
*\*Corresponding authors: lng@boulder.nist.gov, sdiddams@boulder.nist.gov*

We present and characterize a 2-D imaging spectrometer based on a virtually-imaged phased array (VIPA) disperser for rapid, high-resolution molecular detection using mid-infrared (MIR) frequency combs at 3.1 and 3.8 µm. We demonstrate detection of $CH_4$ at 3.1 µm with >3750 resolution elements spanning >80 nm with ~600 MHz resolution in a <10 µs acquisition time. In addition to broadband detection, rapid, time-resolved single-image detection is demonstrated by capturing dynamic concentration changes of $CH_4$ at a rate of ~375 frames per second. Changes in absorption above the noise floor of $5\times10^{-4}$ are readily detected on the millisecond time scale, leading to important future applications such as real time monitoring of trace gas concentrations and detection of reactive intermediates.
*OCIS codes: 120.6200, 300.6340*

Mid-infrared (MIR) frequency comb spectroscopy is a topic of growing interest with applications to challenging measurement fields such as environmental monitoring, health, security, and chemical diagnostics. Thus far, MIR comb spectroscopic approaches have employed a single point detector [1-3] or upconversion to the near infrared for CCD-based detection [4]. Single point detection has the benefit of simplicity, excellent spectral resolution and high acquisition speed but can suffer from detector saturation and related signal-to-noise ratio limitations [5]. On the other hand, a detector array that distributes the signal among multiple detection channels alleviates saturation while maintaining spectral bandwidth and speed. Here we introduce a spectrometer based on a virtually-imaged phased array (VIPA) disperser [6-8] which operates with an InSb array detector in the MIR. We characterize the resolution, sensitivity, and noise properties of this novel spectrometer and analyze its potential for time-resolved, broad-bandwidth trace gas detection.

Fig. 1(a) outlines the experimental setup for which a 100 MHz femtosecond optical parametric oscillator (OPO) provides broadband light tunable from 2.5 to 4.5 µm [9]. The interference of the pump, idler, and signal provide a beat note which is used to stabilize the OPO cavity length to an rf reference via a PZT-actuated mirror (see Ref [9], Fig. 4(b)). Additionally, a tunable sub-MHz linewidth continuous-wave (CW) source based on difference-frequency generation (DFG) allows measurement of the resolution and the free spectral range (FSR) of the VIPA spectrometer. The VIPA itself is a tilted etalon disperser [6] fabricated from Si (0.8 mm thick, n~3.43 @ 3.3 μm). Except for an AR-coated entrance window, the input side is coated with a high reflector (~99.8%), while the output side is coated with a partial reflector (97-98%). All coatings operate over 2.7 to 4.3 μm. To separate each diffraction order, the VIPA output is cross-dispersed horizontally with a 450 l/mm grating. This provides a 2-D image that is recorded with a liquid-$N_2$ cooled InSb array detector (640×512 pixels, 20 μm pitch, 120 Hz maximum full frame rate). As shown in Fig. 2(b), the image of the broadband OPO light with unresolved 100 MHz comb line spacing appears on the detector array as vertical lines separated by the VIPA free-spectral range (FSR).

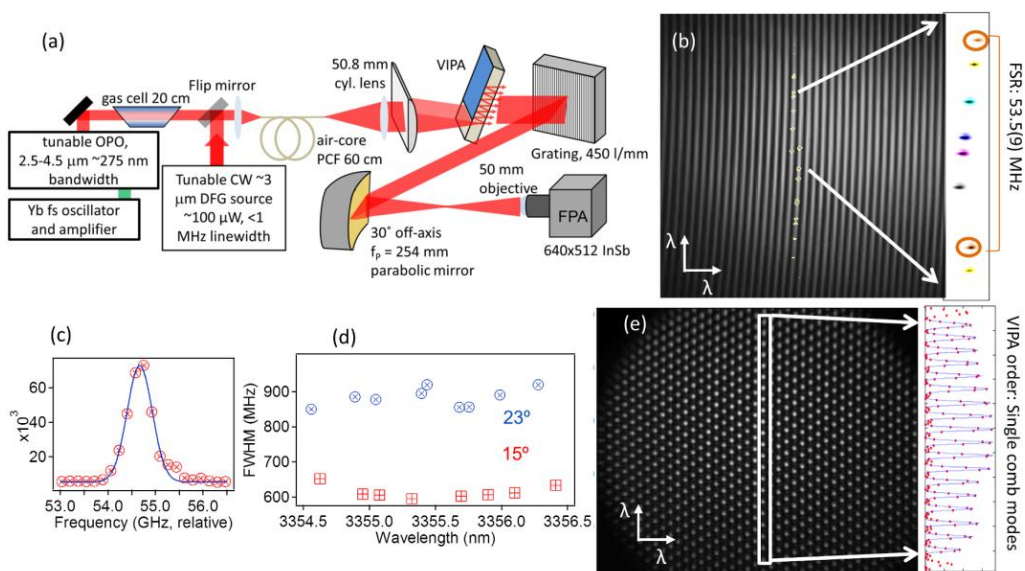

Fig. 1: (a) Experimental setup. MIR light (interchangeable from the OPO or a tunable 3 µm cw source) is coupled into the imaging system through a 60-cm piece of air-core PCF [10]. (b) Image of the broadband 100 MHz, 3 µm comb (gray) through the 97% reflectivity VIPA spectrometer (<10 µs integration time); superimposed is a series of CW frequencies (appearing as spots) that are used to measure the resolution and FSR of the VIPA (right: zoom in of CW spots). (c) Single lineout of a CW spot showing the 600 MHz spectrometer resolution. (d) Spectral resolution for different CW laser frequencies at 15° (squares) and 23° (circles) incidence angles into the VIPA. (e) 2.0 GHz MIR frequency comb at 3.8 µm showing individual comb lines, with a resolution across the image of 600 MHz at a 15° angle of incidence, using a 98% reflectivity VIPA.

We characterized the MIR VIPA in two different ways. First, we used the tunable CW source [Fig. 1(b)] to directly measure the resolution of the 97% reflectivity VIPA imaging system to be ~600 MHz (0.020 cm$^{-1}$). The resolution is roughly constant across the image but depends on the angle of incidence into the VIPA, as is demonstrated in Fig. 1(d). The FSR of the VIPA is experimentally determined as 53.5(9) GHz, which agrees with the expected value of 54.5(2) GHz at a 15° incidence angle. Secondly, these results are confirmed in a separate VIPA spectrometer experiment using a 2.0 GHz MIR comb at 3.8 μm (generated from a cavity-filtered 137 MHz OPO) that allows individual comb lines to be resolved [Fig. 1(e)]. This second VIPA, having a nominal 98% reflectivity and a 15° angle of incidence, also yields a VIPA-system resolution of 600 MHz. Frequency-scanning such a mode-resolved comb will permit MIR spectroscopy at a resolution given by the comb linewidth instead of the VIPA [7,11].

In order to detect trace amounts of a sample of interest, it is important to understand the noise limits of the spectrometer. Figure 2 summarizes our analysis of the noise and stability of subsequent averaging of a 640×512 image of broadband MIR light centered at ~3.1 μm. A data set consisting of 1500 sequential VIPA images of the OPO spectrum was collected with 35 μs integration time and 8.3 ms readout time per image. The images were grouped into blocks of averages (1, 2, 5, 10, 25, etc.) and we calculate $\ln(S_1/S_2)$ to quantify the noise between subsequent images $S_1$ and $S_2$. The quotient $S_1/S_2$ is approximately equal to 1+ε (where |ε|<<1) such that $\ln(S_1/S_2)$ = ε, which we refer to as the effective absorption noise floor. Fig. 2 is a plot of the standard deviation of ε as a function of $N$ averages. Notably, we verified that the fluctuation in the noise floor of ~5×10$^{-4}$ between single images is near the detector shot noise limit. This noise floor decreases as the square root of N, but quickly levels off after 5 averages at an absorption value of ~2×10$^{-4}$. We attribute this behavior to 1/f noise (flicker) of the output power of the laser system, which was confirmed by an independent measurement of the amplitude noise of the MIR light. Furthermore, illuminating only the camera with a thermal source (flashlight) of comparable MIR power revealed that the InSb array itself was not responsible for the flicker noise floor. In this case, we observe continued square root of N averaging for an additional decade.

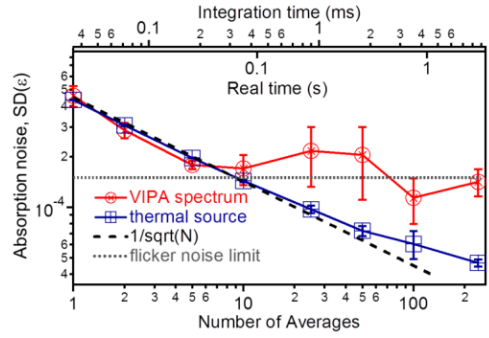

Fig. 2: Noise characteristics of the MIR VIPA spectrometer obtained from a series of 1500 images. The standard deviation (SD) of the absorption noise, ε, is plotted versus the number of spectra averaged. Top axes indicate the exposure (integration) time and the real time required to read out the image. The error bars indicate a repetition of the measurement on separate collections of images.

Quantifying the absorption noise floor allows us to predict the minimum detectable absorption coefficient $\alpha_{min}$ of a gas sample when integrated over a given path length. For example, with a 200 m multipass cell, the noise level of ~2×10$^{-4}$ in a 42 ms data readout time (five averages) would correspond to a sensitivity of $\alpha_{min} \approx 1\times 10^{-8}$ cm$^{-1}$. The sensitivity could be further improved with the longer interaction length provided by a high-finesse enhancement cavity [12].

Fig. 3 shows results of preliminary spectroscopic measurements. The average of 10 spectra (7 μs integration time each) is recorded from a 20 cm sample cell containing a mixture of methane gas (<1%) in $N_2$ (300 Torr total pressure). The cell is subsequently evacuated, and the process is repeated to obtain a background spectrum. Fig. 3(a) displays the resulting 2-D spectrum after the background VIPA image is divided by the sample image. Due to the finite size of the InSb array, the

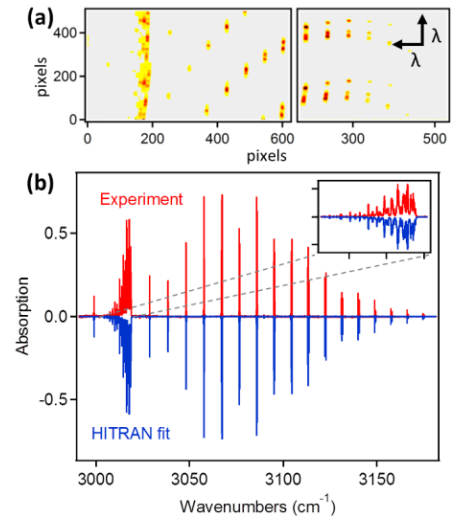

Fig.3 (a) Sample VIPA image (background/signal) for a 20 cm cell with <1% $CH_4$ in 300 Torr $N_2$. Images with two grating angles are concatenated to cover ~210 nm of the OPO bandwidth. (b) Experimental lineout spectrum (red, top). The blue (bottom) spectrum is a fit to the experimental data based on HITRAN line positions and strengths.

grating in the VIPA spectrometer is rotated to obtain a second set of spectra to provide coverage across ~210 nm of the OPO bandwidth.

The VIPA FSR is evident in the 2-D spectrum as the repeating pattern of peaks from top to bottom. This information is used to provide an initial frequency axis for the VIPA spectrum. Each VIPA stripe is linearized (pixel by pixel, summing over 5 pixels perpendicular to the stripe, given by the 1/e width of the stripe) and then the stripes are concatenated end-to-end (after one FSR) to construct the traditional lineout spectrum [top red curve in Fig. 3(b)]. The bottom blue curve in Fig. 3(b) is a fit to the measurement based on peaks in the HITRAN database. From each HITRAN line a Voigt profile is constructed with three fit parameters: center frequency, amplitude, and Lorentzian width (with a fixed Gaussian width consisting of the Doppler width convolved with the instrument response). In this case the Lorentzian width (due to pressure broadening) dominates the lineshape. The fit returned a partial methane pressure of ~1.2 Torr, with residuals across the full spectrum in Fig 3(b) of RMS = 0.0017. This value is presently dominated by uncertainties in the absorption cross-section of the gas, the instrumental line shape response, unaccounted impurities in the sample, and the convergence of the multi-parameter fitting algorithm itself. These issues will be important for more quantitative VIPA-based spectroscopy and will be addressed in future work.

Finally, the rapid and broadband detection capability of a VIPA spectrometer also enables time-resolved measurements covering many spectral lines at the same time [13]. As a demonstration, we capture spectra at a rate of ~375 frames/s (320×320 pixels, 50 cm$^{-1}$ bandwidth) as methane fills an evacuated 20 cm cell. Fig 4(a) shows a subset of the methane lines, and Fig. 4(b) focuses in on the time evolution of a single set of lines using a false-color intensity plot. Displayed in this way, one can see the absorption of each peak grow and broaden as the pressure increases to ~3 Torr. Fig. 4(c) shows a fixed-wavenumber slice centered on a single peak, demonstrating the rise from zero absorption to the full peak height.

To evaluate the sensitivity of the VIPA spectrometer for dynamic spectroscopy, the difference between vertical time slices through the 3D plot was calculated. These differences provide information on the absorption change between frames. In Fig. 4(d) the bottom red curve displays the difference between two subsequent spectra (Δt =2.67 ms). The main peak is readily detected, with a signal-to-noise of ~5 at this data capture rate. The top blue curve is the difference of two images separated by 5.34 ms, where the S/N of the main peak is now ~10 and the side peaks are clearly resolved above the noise. We confirm that the noise level in this type of analysis is similar to the first point on Fig. 2, with subtraction of single images displaying absorption noise on the order of 4.5×10$^{-4}$. Beyond trace gas detection, wide-bandwidth time-resolved spectroscopy and chemical dynamics are additional potential applications of this system.

We thank K. Knabe, P. Del' Haye, A. Zolot, and N. Newbury for their contributions and F. Benabid for providing the air core PCF. This research was funded by NIST, NSF, NRC, and DHS. Any mention of commercial products does not constitute an endorsement by NIST.

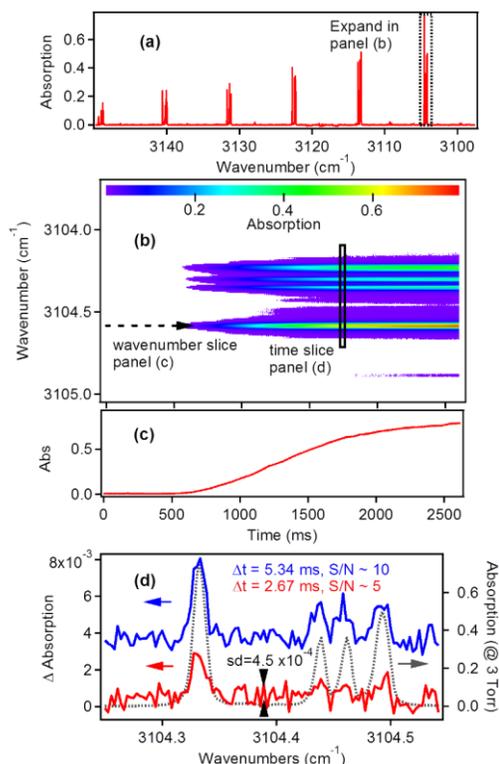

Fig.4 (a) A lineout spectrum from a single frame (320x320 pixels) of a 1500 frame movie of methane gas entering a 20 cm gas cell (b) Zooming in to one set of methane lines, a false color plot is constructed showing the absorption increasing with time at a rate of 2.67 ms/frame. (c) The absorption of a single methane line is monitored versus time. (d) The difference (Δ) of absorption spectra from the VIPA image highlighting the achievable detection sensitivity between 2 images separated by 2.67 ms (red, bottom) and 5.34 ms (blue, top); overlaid is the 3 Torr methane spectrum (dashed).

**References**
1. F. Adler, P. Masłowski, A. Foltynowicz, K. C. Cossel, T. C. Briles, I. Hartl, and J. Ye, Opt. Express **18**, 21861 (2010).
2. E. Baumann, F. R. Giorgetta, W. C. Swann, A. M. Zolot, I. Coddington and N. R. Newbury, Phys. Rev. A **84**, 062513 (2011).
3. B. Bernhardt, E. Sorokin, P. Jacquet, R. Thon, T. Becker, I. T. Sorokina, N. Picqué, and T. W. Hänsch, Appl. Phys. B **100**, 3 (2010).
4. T. A. Johnson and S. A. Diddams, Appl. Phys. B **107**, 31 (2012).
5. N. R. Newbury, I. Coddington, and W. C. Swann, Opt. Express **18**, 7929 (2010).
6. M. Shirasaki, Opt. Lett. **21**, 366 (1996).
7. S. A. Diddams, L. Hollberg, and V. Mbele, Nature **445**, 627 (2007).
8. M. J. Thorpe, D. Balslev-Clausen, M. S. Kirchner, and J. Ye, Opt. Express **16**, 2387 (2008).
9. F. Adler, K. C. Cossel, M. J. Thorpe, I. Hartl, M. E. Fermann and J. Ye, Opt. Lett. **34**, 1330 (2009).
10. F. Benabid, Phil. Trans. R. Soc. A **364**, 3439 (2006).
11. L. C. Sinclair, K. C. Cossel, T. Coffey, J. Ye, and E. A. Cornell, Phys. Rev. Lett. **107**, 093002 (2011).
12. A. Foltynowicz, P. Masłowski, A. J. Fleisher, B. Bjork, and J. Ye, Appl. Phys. B, DOI: 10.1007/s00340-012-5024-7 (2012).
13. M. J. Thorpe, F. Adler, K. C. Cossel, M. H. G. de Miranda, and J. Ye, Chem. Phys. Lett. **468**, 1 (2009).

1. F. Adler, P. Masłowski, A. Foltynowicz, K. C. Cossel, T. C. Briles, I. Hartl, and J. Ye, "Mid-infrared Fourier transform spectroscopy with a broadband frequency comb," Opt. Express **18**, 21861-21872 (2010).
2. E. Baumann, F. R. Giorgetta, W. C. Swann, A. M. Zolot, I. Coddington, and N. R. Newbury, "Spectroscopy of the methane $\nu_3$ band with an accurate mid infrared coherent dual-comb spectrometer," Phys. Rev. A **84,** 062513 (2011)
3. B. Bernhardt, E. Sorokin, P. Jacquet, R. Thon, T. Becker, I. T. Sorokina, N. Picqué, and T. W. Hänsch, "Mid-infrared dual-comb spectroscopy with 2.4 µm $Cr^{2+}$:ZnSe femtosecond lasers," Appl. Phys. B **100**, 3 (2010)
4. T. A. Johnson, S. A. Diddams, Mid-infrared upconversion spectroscopy based on a Yb:fiberfemtosecond laser", Appl. Phys. B., **107**, 31 (2012).
5. N. R. Newbury, I. Coddington, and W. C. Swann, "Sensitivity of coherent dual-comb spectroscopy," Opt. Expr. **18**, 7929 -7945 (2010).
6. M. Shirasaki, "Large angular dispersion by a virtually imaged phased array and its application to a wavelength demultiplexer," Opt. Lett. **21**, 366-368 (1996).
7. S. A. Diddams, L. Hollberg, and V. Mbele, "Molecular fingerprinting with the resolved modes of a femtosecond laser frequency comb," Nature **445**, 627-630 (2007).
8. M. J. Thorpe, D. Balslev-Clausen, M. S. Kirchner, and J. Ye, "Cavity-enhanced optical frequency comb spectroscopy: application to human breath analysis," Opt. Expr. **16**, 2387-2397 (2008).
9. F. Adler, K. Cossel, M. Thorpe, I. Hartl, M. Fermann and J. Ye, Opt. Lett. **34**, 1330 (2009).
10. The air core PCF is generously provided by Fetah Benabid at Xlim, CNRS UMR, France. See for example, F. Benabid, "Hollow-core photonic bandgap fibre: new light guidance for new science and technology," Phil. Trans. R. Soc. A **364**, 3439-3462 (2006).
11. L. C. Sinclair, K. C. Cossel, T. Coffey, J. Ye, and E. A. Cornell, "Frequency comb velocity-modulation spectroscopy," Phys. Rev. Lett. **107**, 093002 (2011).
12. A. Foltynowicz, P. Masłowski, A. J. Fleisher, B. Bjork, and J. Ye, "Cavity-enhanced optical frequency comb spectroscopy in the mid-infrared - application to trace detection of $H_2O_2$," DOI: 10.1007/s00340-012-5024-7 (2012).
13. M. J. Thorpe, F. Adler, K. C. Cossel, M. H. G. de Miranda, and J. Ye, "Tomography of a supersonically cooled molecular jet using cavity-enhanced direct frequency comb spectroscopy," Chem. Phys. Lett. **468**, 1-8 (2009).